\documentclass[aps,prc,twocolumn,groupaddress,showkeys]{revtex4-1}
\pdfoutput=1
\usepackage{graphicx} 
\usepackage{dcolumn} 
\usepackage{bm}
\usepackage{color}
\usepackage{hyperref}
\usepackage{graphicx}
\usepackage{amsmath}

\begin{document}

\title{Taming nucleon density distributions with deep neural network method}
\author{Zu-Xing Yang$^{1,2}$}
\email[]{yangzuxing16@impcas.ac.cn}
\author{Wei Zuo$^{1,2}$}
\email[]{zuowei@impcas.ac.cn}
\author{Peng Yin$^{1,3}$}
\author{Xiao-Hua Fan$^{4}$}
\affiliation{$^1$Institute of Modern Physics, Chinese Academy of Sciences, Lanzhou 730000, China\\
$^2$School of Nuclear Science and Technology, University of Chinese Academy of Sciences, Beijing 100049, China\\
$^3$Department of Physics and Astronomy, Iowa State University, Ames, IA 50011, USA\\
$^4$School of Physical Science and Technology, Southwest University, Chongqing 400715, China
}

\begin{abstract}
We investigate the density distributions of finite nuclei employing a well-designed deep neural network method. 
We calculate the target nucleon density distributions with Skyrme density functional theories, which are used to train the networks.
We find that the training with only about $10\%$ nuclei ($300-400$) is sufficient to  describe the nucleon density distributions of all the nuclear chart within 2\% relative error. 
The relative error comes to 5\% when about 200 proton(neutron) density distributions are used for training.
We obtained very similar results for different Skyrme density functional theories. 
Therefore the ability to train networks is weakly dependent on the theoretical model.
Moreover, in the process of machine learning, there is a turning point showing the transition from the Fermi-like distribution to the realistic Skyrme distribution, which provides significant properties of convergence process.

\end{abstract}

\maketitle

\section{Introduction}

Since the discovery of the atomic nucleus\cite{EXP00}, density information has played a crucial role in the study of nuclear structure, especially in explaining the nature of nuclear force and the fundamental properties of nuclear matter\cite{EXP01}. 
For example, in studies of high-momentum tails(HMT) induced by short-range correlations(SRC), it has been found that the percentage of tails is highly dependent on the density\cite{SRC,Yin:2013cc,Yin:2013hwa}. 
The asymmetry potential is density-dependent, which is influenced by the Pauli blocking of the $\Delta$-decay and the thickness of the neutron skin\cite{SE}. 
Moreover, The equation of state(EoS) of nuclear matter is used to describe the  neutron stars(NS) mergers\cite{AP1}.

In the 1950s, Hofstadter first measured the charge density of protons using electron scattering experiments and described the density distribution of some nuclei on that basis\cite{EXP02}. 
Whereafter, two-parameter Fermi (2pF) / three-parameter Fermi (3pF) distributions\cite{DES00,DES01,DES02} or Fourier-Bessel expansions\cite{DES02, DES03} have been used to roughly describe the shape of nuclear density profile. 
The density functional theory (DFTs), e.g., Skyrme-Hartree-Fock (SHF) method, is one of the most efficient and widely used models to study the bulk properties of nuclei, especially the density distribution\cite{SHF00, SHF01, SHF02}.
With the creation of the Bardeen-Cooper-Schrieffer(BCS) theory\cite{BCS00}, the understanding of superconductivity theory has been subverted and new blood has been injected into nuclear physics. 
Bogoliubov imported the generalized Bogoliubov transform to include both the particle-hole and particle-particle parts of nuclear force\cite{BCS01}, which, of course, made for a more refined description of density. 
Currently, almost all models describing nuclear structure can calculate density distributions.
In 2016, density information for the short-lived nucleus $^{34}$Si was also obtained by comparing the experimental measurements for the removal of an $l = 0$ proton with the results of reaction theories \cite{EXP05}.  
Some simulations, based on transport models, also have a positive effect on the discovery of density profile. 
For example, using density profile as a initialization-condition, the bubble structure was researched by heavy ion collisions(HIC) simulation of p+$^{48}$Si\cite{BUU00}. 

Back propagation neural network (BPNN) method\cite{AI07, AI08, AI09, AI10, AI11, AI12, AI13, AI14}, which is the most popular and powerful machine learning tools, has been remarkablly developed in computer vision (CV)\cite{AI00, AI01} and natural language processing (NLP)\cite{AI02, AI03} in recent years. 
In the application to the nuclear physics, it is also becoming prevailing. 
By training neural networks with the X-ray radiation from neutron stars,  the relationship between pressure and mass density has been estimated\cite{AI04}. 
Combining the Hartree-Fock-Bogoliubov(HFB) and multilayer neural network methods, the  nuclear complexity was tamed, where excited state energy, nuclear deformation etc. of nuclei are predicted\cite{AI05}.  
Excellent results have been obtained using Bayesian neural network (BNN) methods for the extrapolation of binding energies from stable nuclei to drip-line region\cite{AI06}.
The artificial neural network method was used to extrapolate the ground state properties of nuclei calculated by {\it ab-initio} approach to large basis space\cite{Negoita:2018yok,Negoita:2018kgi}. 
Until recently, Google proposed a hybrid quantum-classical machine learning model for training beyond classical data types\cite{AIquantum}.
Inspired by these successful applications of the machine learning method, we briefly modify the traditional DNN model by using the determined mass number as a physical condition.

\section{The neural network method and Machine learning process}


\begin{figure}
\includegraphics[width=8 cm]{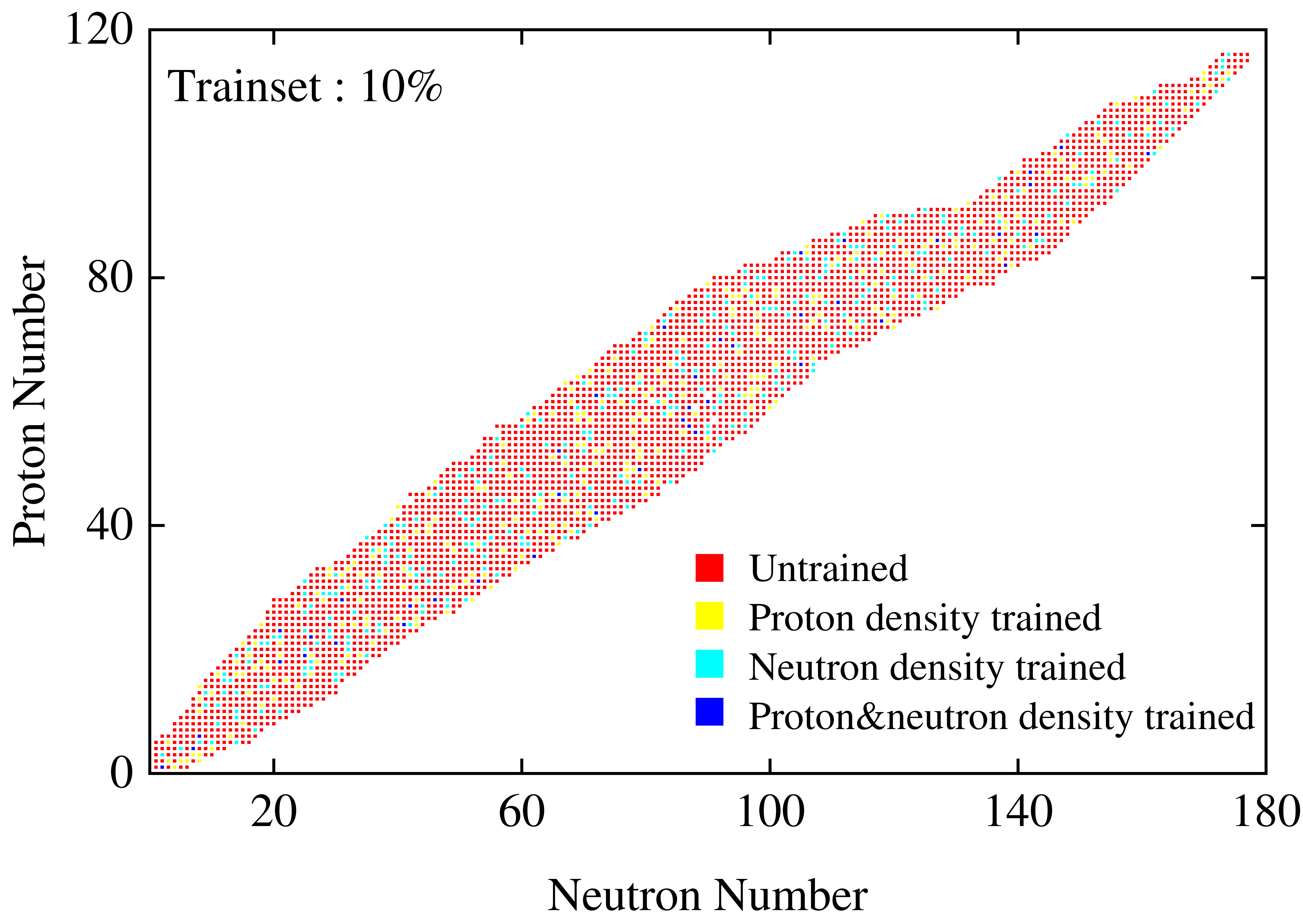}
\caption{\label{fig:nuclearchart}(Color online) Position of the trained nuclide in the nuclear chart with 10\% of dataset as train set.}
\end{figure}

This work is aimed at mapping out the nucleon density distributions of almost all the nuclei employing the neural network method by training networks with density distributions of very limited nuclei.
This is a logistic regression problem in supervised learning, which statistically can be seen as a maximum likelihood estimation (MLE)\cite{AIweight},

\begin{equation}
\begin{aligned} \mathbf{w}^{\mathrm{MLE}} &=\arg \max _{\mathbf{w}} \log P(\mathcal{D} | \mathbf{w}) \\ &=\arg \max _{\mathbf{w}} \sum_{i} \log P\left(\mathbf{y}_{i} | \mathbf{x}_{i}, \mathbf{w}\right), \end{aligned}
\label{eq_Wmle}
\end{equation}
where $\boldsymbol{\rm{w}}$ and  $\boldsymbol{\rm{w^{MLE}}}$ denote the weights of neural networks during learning and the weights of neural networks where the mapping is implemented with the greatest probability, respectively. 
$\boldsymbol{\rm{x}_i}$ and $\boldsymbol{\rm{y}_i}$ represent the input and output, respectively. 
In this work, the input is $\lbrace Z_{num} , N_{num} , \tau \rbrace $ where $Z_{num}$ and $N_{num}$ denote the proton and neutron numbers, respectively. 
$\tau$ corresponds to the proton ($\tau = 1$) or neutron ($\tau = 2$). 
The output is the nucleon ($\tau=1$ for proton and $\tau=2$ for neutron) density distribution of the nucleus ($Z_{num}$, $N_{num}$). 
That is, the neural network is to produce the proton or neutron density distribution of a nucleus for a given set of $\lbrace Z_{num} , N_{num} , \tau \rbrace $ .

BPNN, which was proposed by Rumelhart and McClelland in 1986, is adopted to achieve Eq.(\ref{eq_Wmle}). 
The method adjusts multi-layer feedforward neural network with back propagation algorithm of error, by which the weight of the link chain converges toward its stated goal $\rm{w^{MLE}}$.

$network~structure$ Neural networks typically contain a large number of interrelated neurons in the input layer, output layer, and several hidden layers. 
We used about 860,000 parameters located in eight layers ($l\in[1,8]$),
\begin{equation}\begin{aligned}
z_{i}^{(l+1)} &=b_{i}^{(l)}+\sum_{j=1}^{S_{l}} w_{i j}^{(l)} a_{j}^{(l)} \\
a_{i}^{(l)} &=g\left(z_{i}^{(l)}\right) ,\\
\label{eq_NN}
\end{aligned}\end{equation} 
where $a_{j}^{(l)}$ and $z_{i}^{(l+1)}$ denote the input and output of the $l$-th layer; $w_{i j}^{(l)}$ and $b_{i}^{(l)}$ are weight and bias of the $l$-th layer.
In the input layer and each hidden layer($l\in[1,7]$), the nonlinear activation function $g(x)={\rm{ReLU}}(x)=\max (0, x)$, which has been shown to have extraordinary effects\cite{AI18, AI19}, is employed. 
In the output layer($l=8$), the activation function $g(x)={\rm{Sigmoid}}(x)=\frac{1}{1+e^{-x}}$ is adopted to smooth the results.

The loss function is a key component in network regulation. The mean squared error (MSE) is widely used for regression problems, which can be obtained naturally by taking a normal distribution for $P(\mathcal{D} | \mathbf{w})$  in eq.(\ref{eq_Wmle}). 
However, a physical constraint $\int_{0}^{\infty} 4 \pi \rho(r) r^{2} d r=X$ has to be taken into account in this work, with $X=Z_{num}$ or $N_{num}$. $\rho(r)$ denotes the corresponding nucleon density distribution. 
We therefore introduce a correction to the MSE and define the normalized mean square error(NMSE) as follows,

\begin{equation}
\begin{aligned}
NMSE&=\sum_{\alpha=1}^{M}\frac{1}{N} \sum_{i=1}^{N}\left(\lambda\rho^\alpha_{pre}(r_i)-\rho^\alpha_{tar}(r_i)\right)^{2},\\
\lambda&=\frac{X}{\int_{0}^{\infty} 4 \pi \rho_{pre}(r) r^{2} d r},
\end{aligned}
\label{eq_MSE}
\end{equation}
where $\alpha$ runs over all the nuclei which are adopted to train the networks. $M$ corresponds to the number of nuclei in a training set. 
$i$ runs over all the radial grids for a specified nucleus $\alpha$. In this work we consider the radial range from $r=0$ fm to $r=15$ fm and divide  $r$ into $N=150$ grids for all the nuclei.
$\rho_{pre} = a_{i}^{(8)}$ refers to the nucleon density distribution of a nucleus obtained with the current neural network method during the learning process, and $\rho_{tar}$ denotes the target nucleon density distribution of the same nucleus, which is calculated with the Skyrme DFTs. 
With the correction in Eq.(\ref{eq_MSE}), absolute density becomes irrelevant. 
The conserved quantity allows our networks to be more targeted in capturing scale Features.
The networks hence can be well-converged even with very limited training dataset.

 \begin{figure}
\includegraphics[width=9 cm]{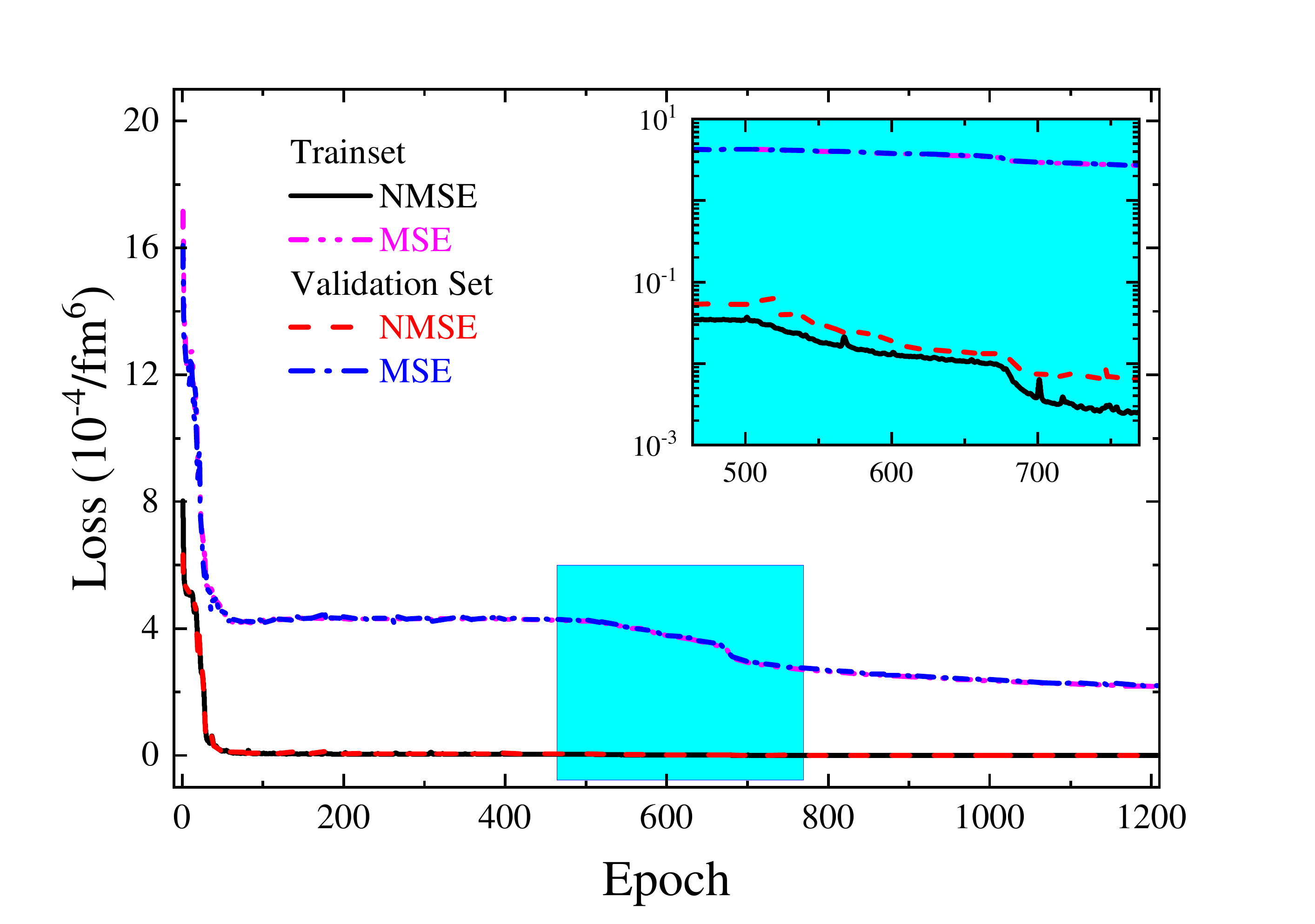}
\caption{\label{fig:10data}(Color online) Loss function NMSE and MSE as a function of epochs, on the training set and the validation set, respectively.
  }
\end{figure}

Combining Eq.(\ref{eq_MSE}) with Eq.(\ref{eq_Wmle}), mapping out the density distributions of a set of nuclei for a training set is equivalent to solving the following equation, 
\begin{equation}
\begin{aligned} 
\mathbf{w}^{\mathrm{MLE}} &=\arg \min _{\mathbf{w}} NMSE(\rm{w}).
\end{aligned}
\label{eq_Wmle2}
\end{equation}
During the machine learning, we employ  the adaptive moment estimation(Adam)\cite{AI15} for the optimizer to operate gradient descent.
We manipulate the learning rate according to the variation in error at different epochs. 
We take the Keras with a Tensorflow backend \cite{AI16} to train the networks. The training results become well converged within 2000 epochs. 
One epoch represents that all the neurons have been trained once.

$training~process$ About 3400 nuclei have been discovered in laboratories to date.
Various theoretical approaches have been used to calculate the nucleon density distributions of these nuclei. 
In this section, the target nucleon density distributions are obtained by the SHF+BCS approxiamation with the SKM* interaction, which we refer to as SKM*-SHF.
For a given training set, we take a fraction of the nucleon density distributions of these nuclei to train the neural networks.  
We calculate density using the following equation\cite{GSHF}:

\begin{equation} 
\rho_{\tau}(r)=\sum_{n_{\beta} j_{\beta} l_{\beta}} w_{\beta} \frac{2 j_{\beta}+1}{4 \pi}\left(\frac{R_{\beta}}{r}\right)^{2}
\label{eq_rho}
\end{equation}
where $w_{\beta}$ and $R_{\beta}$ denote the pairing weight and radial wave function for each single particle state, respectively. The total wave function reads:
\begin{equation} 
\varphi_{\beta}(\boldsymbol{r})=\frac{R_{\beta}(r)}{r} \mathcal{Y}_{j_{\beta} l_{\beta} m_{\beta}}(\theta, \phi).
 \label{eq_wfs}
\end{equation} 
The functions $\mathcal{Y}_{j_{\beta} l_{\beta} m_{\beta}}(\theta, \phi)$ are spinor spherical harmonics.

In our first application, we train the networks with $10\%$ of the entire dataset, formed by about 6800 sets of neutron/proton density distributions of about 3400 nuclei. 
We present the locations of corresponding nuclei in Fig.~\ref{fig:nuclearchart}. 
Note that we choose these $10\%$ of the dataset randomly. 
Therefore both the proton and neutron density distributions of some nuclei (blue symbols in Fig.~\ref{fig:nuclearchart}) are adopted in these $10\%$ data, whereas only one of the two nucleon density distributions [either the proton (yellow symbols in Fig.~\ref{fig:nuclearchart}) or neutron density distribution (blue symbols in Fig.~\ref{fig:nuclearchart})] is employed for other adopted nuclei. 
The red symbols in Fig.~\ref{fig:nuclearchart} corresponds to the nuclei which are not used in the training process.


\begin{figure}
\includegraphics[width=9 cm]{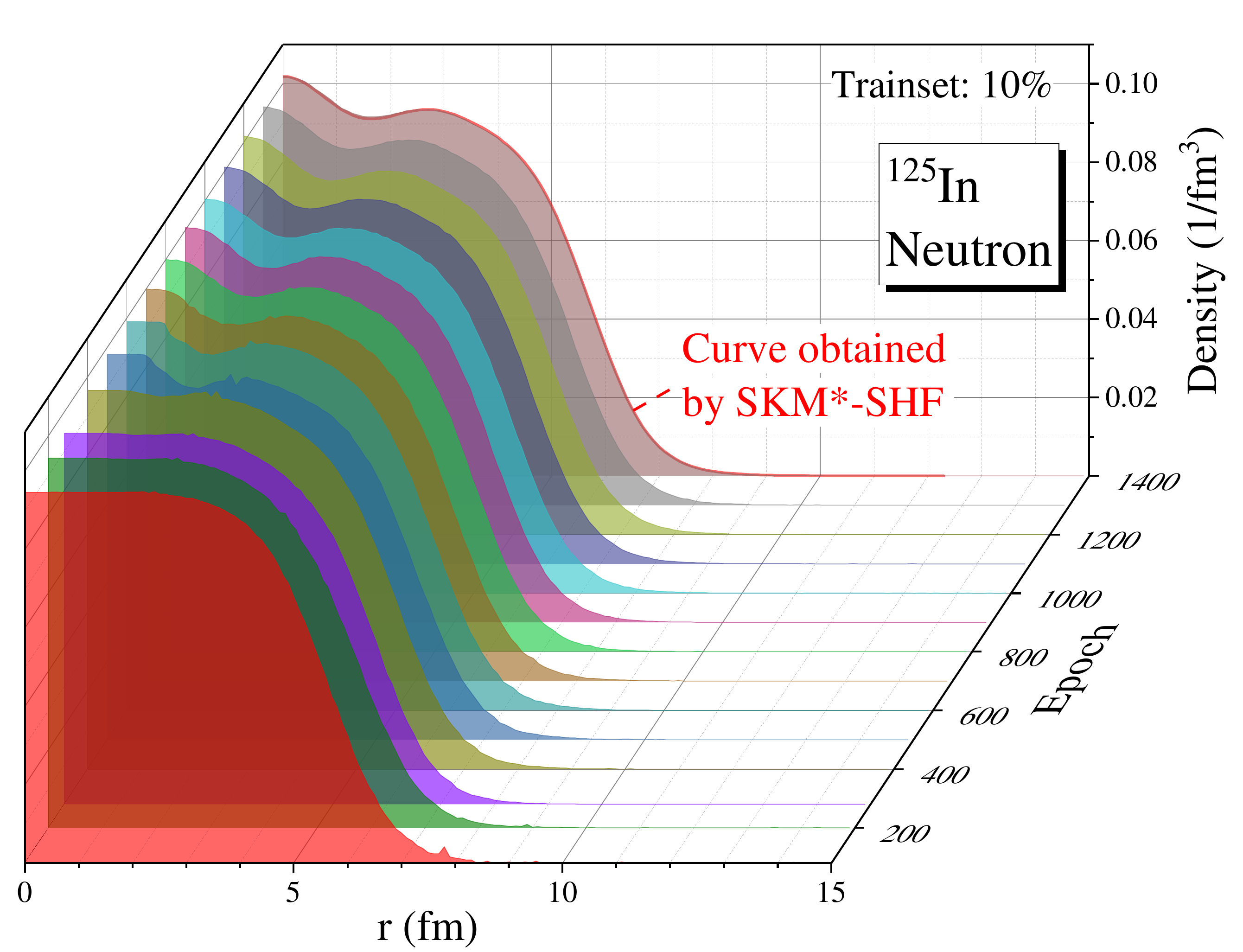}
\caption{\label{fig:10learningpro} (Color online) The predicted neutron density distributions of $\rm{^{125}In}$ as a function of epochs during the training. The red line at 1400th epoch is target distribution calculated by SHF model.
  }
\end{figure}

In Fig.\ref{fig:10data} we present the loss function NMSE and MSE as functions of epoch, for both the training set and the validation set. For the training set, we use  $10\%$ of the entire dataset to train the networks as shown in Fig.~\ref{fig:nuclearchart}. 
We found that the NMSE and MSE decreases with epoch simultaneously during the training process, which indicating that the results tend to be converged with networks trained sufficiently. 
Meanwhile, another $10\%$ of entire dataset was evaluated after each training epoch as the validation test.
It is clear that, both for NMSE and MSE, there is a high consistency of errors on training sets and validation sets. 
This shows that no under-fitting or over-fitting during the training process, which illustrates that the neural network method is able to describe the nucleon density distributions of the entire nuclear chart by training the networks with only $10\%$ of the dataset. 
Note that NMSEs are much smaller than MSEs in Fig.\ref{fig:10data} since we train the networks with the NMSE other than the MSE. 
We therefore treat MSE as an amplifier for NMSE. 
With the help of this amplifier, we notice that the two significant convergence processes emerge during the training process: the first one is at about 50 epochs and the second one is from 500 to 700 epochs.

What do the two convergences represent?
In Fig.\ref{fig:10learningpro} we show the evolution of the neutron density distribution of a random nuclide $^{125}$In on the test set during the training process in Fig.\ref{fig:10data}. 
The first significant convergence brings the density distribution from chaos to order, which is achieved rapidly within around 50 epochs.
The red block at 100th epoch,  which is Fermi-like, is an exemplary distribution after the first significant convergence. 
The second significant convergence emerges from about 500th epoch, after which the Fermi-like distribution evolves to that calculated by the SKM*-SHF method.
In order to make the value of the loss function as small as possible, the network naturally gets the Fermi-like distribution, which is like the process of human understanding of nuclear density profile in the past 70 years, from naive to refined. 
Epoch 500 is a crossover point, which is the starting point where the loss function falls again after a long period of steady state.
We therefore refer to the 500th epoch as a wisdom-point (W-point). 

Regarding the W-point, here are some interesting conclusions: 

i. the W-point appears earlier (later) as the amount of training data increases (decreases);

ii. The W-point does not appear if the training data is too small;

iii. The appearance of the W-point is due to the particle number condition we added to the loss function.

After the second convergence, the model is still fine-tuned. 
Until the 1400th epoch, the distribution is almost the same as the target distribution, which implies that the neural network method is rather useful for the current application. 
We retain the neural network within 2000 epochs where the validation set error is minimum.


\section{Results and discussions}

\begin{figure}
\includegraphics[width=9 cm]{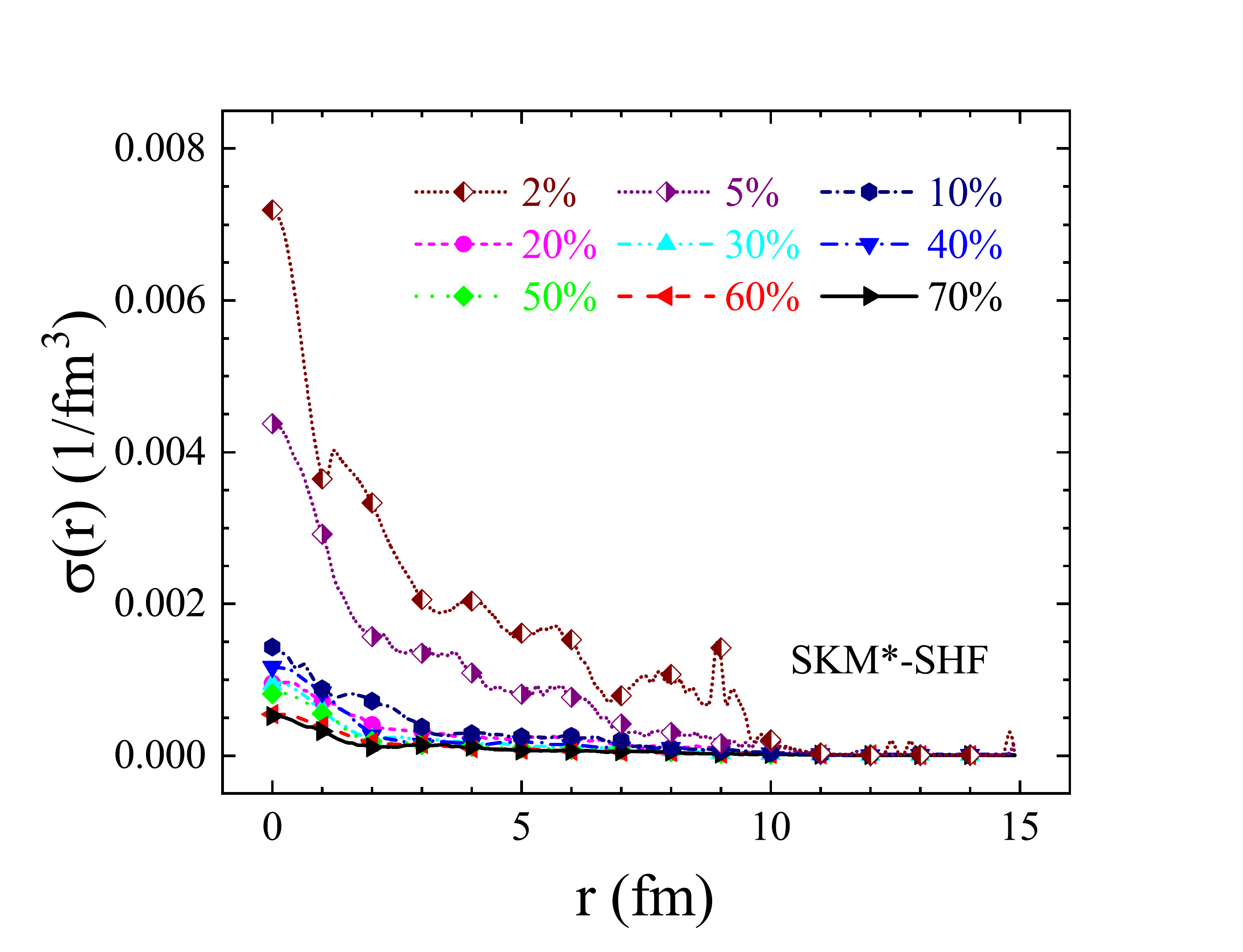}
\caption{\label{fig:meanlossr} (Color online) The error $\sigma(r)$ of predicted density distributions as a function of radii with different percentages of the dataset as train sets by evaluating 100 random distributions on the validation set.
  }
\end{figure}

\begin{figure*}[htbp] %

  \centering
  \includegraphics[width=1.0\linewidth]{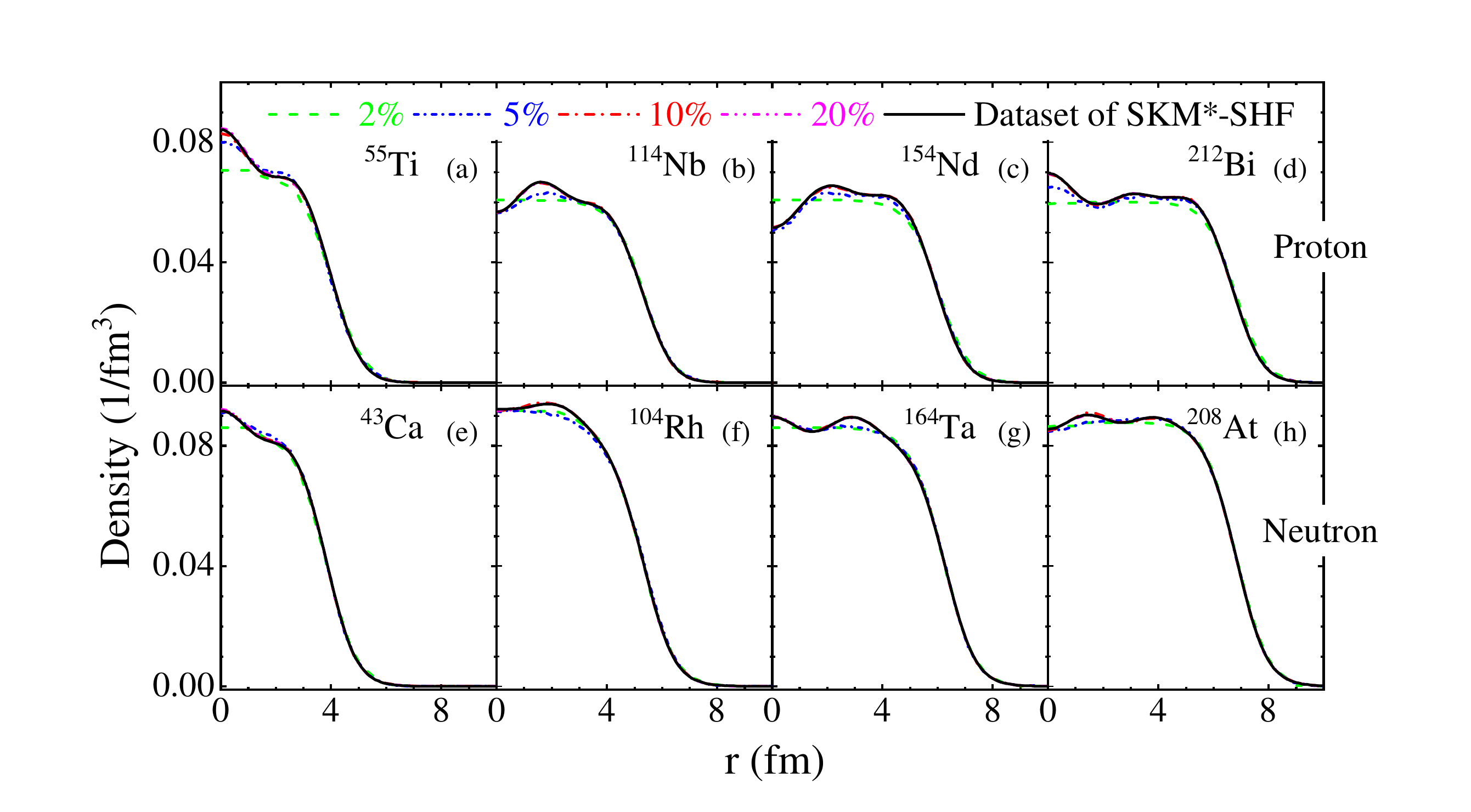}
  \caption{ (Color online) The proton density distribution of four random nuclei-$^{55}$Ti(a), $^{114}$Nb(b), $^{154}$Nd(c), $^{212}$Bi(d) and the neutron density distribution of four random nuclei-$^{43}$Ca(e), $^{104}$Rh(f), $^{164}$Ta(g), $^{208}$At(h) in test set predicted by the trained neural-net with different fractions of dataset, and compared with original data from SKM*-SHF. }\label{fig:example}
\end{figure*}

\begin{figure}
\includegraphics[width=9 cm]{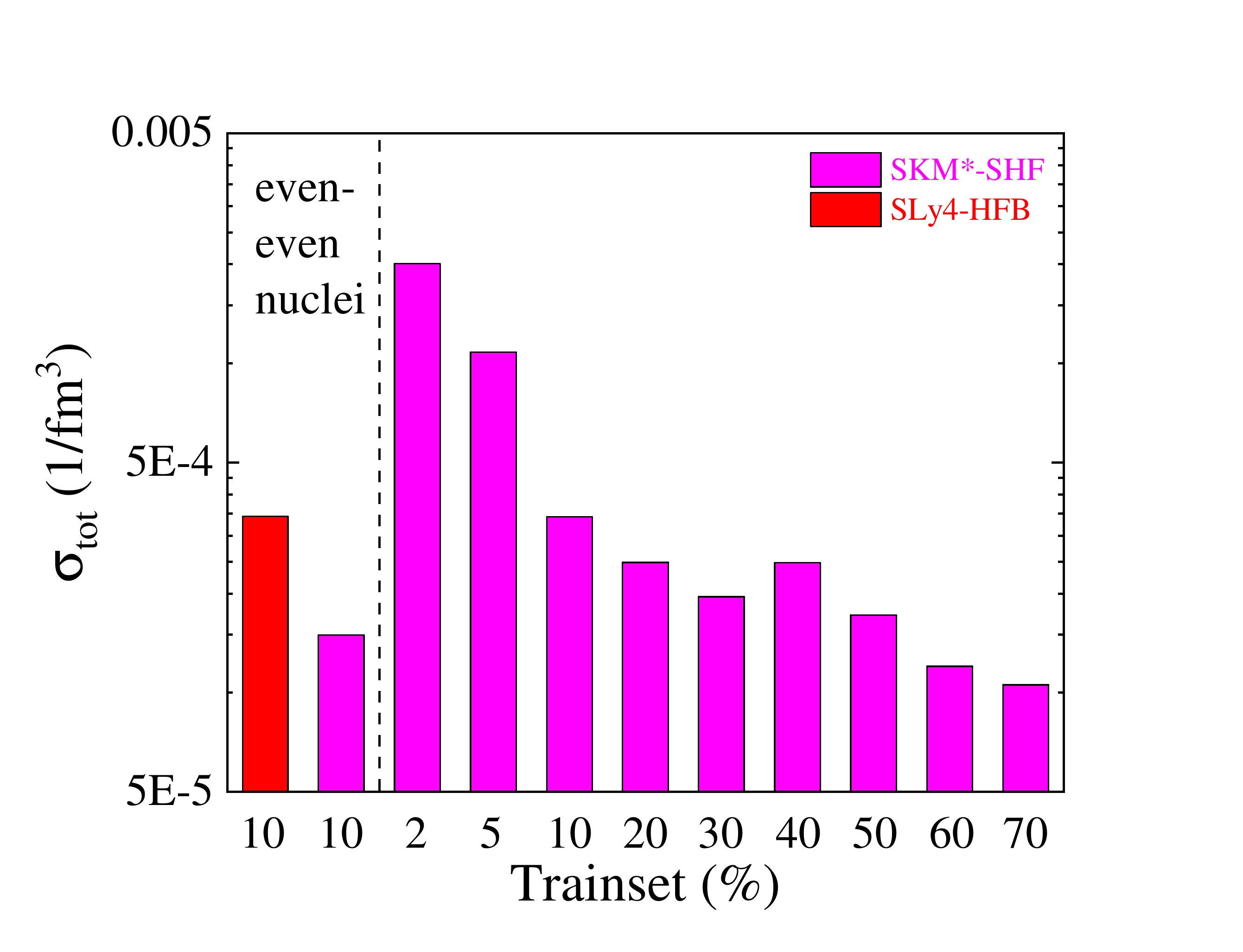}
\caption{\label{fig:lossrate} (Color online) The error $\sigma_{tot}$ of predicted density distributions as a function of  different percentages of the dataset with SKM*-SHF and SLy4-HFB.
  }
\end{figure}

In the following applications, we use two Skyrme DFTs to calculate the target nucleon density distributions in order to investigate the dependence of the neural network predictions on the theoretical model. 
One of the Skyrme DFTs is the SKM*-SHF as we mentioned in the above section. The other one is the Hartree-Fock-Bogoliubov method using the SLy4 interaction, which we refer to as SLy4-HFB.
As for the SKM*-SHF model, nine neural network models with 2\%, 5\% and  10\%-70\% of the dataset adopted to train the networks are used to investigate the convergence pattern.
However, we take 10\% of the dataset, which are all even-even nuclei, to train the networks for the SLy4-HFB model.
Moreover, we use identical inputs for the training processes with SKM*-SHF and SLy4-HFB to guarantee their consistency.

We present in Fig.\ref{fig:meanlossr} the error distribution obtained with the nine neural network models based on the SKM*-SHF data training. 
The error distribution is the standard deviation between the result of the neural network method, $\lambda\rho_{pre}(r)_{i}$ and the result obtained by the SKM*-SHF method, $\rho_{tar}(r)_{i}$, and is written as,
\begin{equation}
\begin{aligned}
\sigma(r)=\sqrt{\frac{\sum_{i=1}^{A}\left(\lambda\rho_{pre}(r)_{i}-\rho_{tar}(r)_{i}\right)^{2}}{A}}.
\end{aligned}
\label{eq_sd}
\end{equation}
The subscript $i$ runs over all $100$ nucleon density distributions ($A=100$) which are randomly selected.
We find in Fig.\ref{fig:meanlossr} that the maximum value of the error distribution is within $0.0015$ $\rm{fm^{-3}}$ for the train set with 10\% dataset. 
Comparing with the saturation density of the proton or neutron, about $0.08$ $\rm{fm^{-3}}$) (half of the nuclear saturation density, $0.16$ $\rm{fm^{-3}}$),  the central relative error (CRE) is less than 2\%. 
For the train set with less than 10\% dataset, we can see the error increases dramatically.
For the case with 5\% dataset, the predicted CRE is between 5\% and 6\%.
However, even with only 2\% of the dataset used for training, the predicted CRE is less than 10\%.
The predicted CRE reduces as more data are included in the train set.

Fig.\ref{fig:example} shows four random  proton density distributions and four random  neutron density distributions predicted by the trained neutral-nets with different fractions-2\%, 5\%, 10\% and 20\% of dataset. 
It is clear that the results predicted by the models trained with 10\% and 20\% of the dataset overlap almost exactly with the curves obtained from SKM*-SHF calculations.
Of particular note is the green curve, which is almost a standard Fermi distribution.
There was no second convergence during the training for this curve, i.e. the W-point did not occur.
This is acceptable because the training data at this point is rather small.
This curve corresponds to a CRE of approximately 10\%.
This value can also be considered as the CRE at the W-point during the training with more than 2\% of dataset.
Naturally, the corresponding error curve(the 2\% one) in Fig.\ref{fig:meanlossr} can also be viewed as the error curve at the  W-point.
The predictions of the model trained with 5\% of the dataset are intermediate between the  2\% one and the  10\% one. 
It has undergone the second significant convergence in part and provides an error of only about 5\%.

In Fig.\ref{fig:lossrate} we present a quantity $\sigma_{tot}$, which represents the normalized total standard deviation and is written as (\ref{eq_Tsd1}),
\begin{equation}
\begin{aligned}
\sigma_{tot}=\sqrt{\frac{\sum_{j=1}^{N}\sigma(r_j)^2}{N}},
\end{aligned}
\label{eq_Tsd1}
\end{equation}
where $N$ denotes the number of radial grids.
For the results in the right part of Fig.\ref{fig:lossrate}, $\sigma_{tot}$ decreases as the amount of training data increase, which is consistent with the results in Fig.\ref{fig:meanlossr}. 


In the left part of Fig.\ref{fig:lossrate} we investigate the model dependence of the current neural network method. 
We randomly selected 340 nucleon density distributions of even-even nuclei to train the networks using the target nucleon density distributions calculated by the SKM*-SHF and SLy4-HFB methods.
We notice that $\sigma_{tot}$ for the results obtained with the SKM*-SHF and SLy4-HFB methods are on the same order of magnitude. 
Therefore the predictive power of the current neural network method is weakly dependent on the theoretical models and interactions.
Comparing the data for 10\% of dataset in the left part of Fig.\ref{fig:lossrate} with that in the right part, we find that $\sigma_{tot}$ for even-even nuclei is smaller than the result which does not distinguish the odd/even status of the proton or neutron numbers.
This can be explained in terms of the complexity of information.
Dataset with only even-even nuclei have lower complexity than the other case.
During the training process, low complexity contains less features, which are easier to captured. 
Therefore, the neural network calculations with only even-even nuclei dataset lead to smaller $\sigma_{tot}$. 

\section{Summary}

In this work we proposed a method to describe the nucleon density distributions of finite nuclei using neural network method. 
Due to the normalized condition in the loss function, the nucleon density distribution was always Fermi-like in the early stage of the machine learning and then tended to the target distribution after the second significant convergence. We therefore defined a W-point at which the second significant convergence emerged. 
We found that the W-point emerged earlier with more data adopted in the training process.
We did not observe the W-point for the machine learning with 2\% of dataset.

We found that the machine learning with 10\% of dataset was sufficient to predict the nucleon density distributions of all the nuclei with high precision. 
The corresponding CRE was less than 2\%, which suggested that the neural network method adopted in this work was rather successful in the current application.
By comparing the results using the target distributions calculated with two different Skyrme DFTs, SKM*-SHF and SLy4-HFB, we found that the predictive power of the neural network was weakly dependent on the adopted theoretical model.

So far, the ground-state density distribution of spherical nuclei has been adequately studied.
Compared to traditional methods, the neural network method not only reduces the complexity of the study, allowing one to avoide complicated many-body problem, but also significantly improves the predictive power with less computational cost.

\section{Acknowledgements}

We acknowledge helpful discussions with Gao-Chan Yong and Jian-Min Dong. This work was supported by the National Natural Science Foundation of
China (Grant Nos. 11435014, 11705240, 11975282) and the 973 Program of China (Grant No. 2013CB834405). This
work was also partially supported by the CUSTIPEN (China-U.S. Theory Institute for Physics with Exotic Nuclei)
funded by the U.S. Department of Energy, office of Science under Grant No. DE-SC0009971.

\end{document}